\documentclass[conference, A4 size (210 x 297 mm)]{IEEEtran}
\IEEEoverridecommandlockouts
\usepackage{cite}
\usepackage{amsmath,amssymb,amsfonts}
\usepackage{algorithmic}
\usepackage{algorithm}
\usepackage{graphicx}
\usepackage{textcomp}
\usepackage{multirow}
\usepackage{xcolor}
\usepackage{lipsum}   
\usepackage{lipsum}
\usepackage[colorlinks=true, citecolor=black, linkcolor=black, urlcolor=black]{hyperref}

\def\BibTeX{{\rm B\kern-.05em{\sc i\kern-.025em b}\kern-.08em
    T\kern-.1667em\lower.7ex\hbox{E}\kern-.125emX}}
    
\begin{document}

\title{Enhancing Wind Power Forecast Precision via Multi-head Attention Transformer: An Investigation
on Single-step and Multi-step Forecasting}

%\and

% \author{\IEEEauthorblockN{Md Rasel Sarkar}
% 	\IEEEauthorblockA{School of Engineering \& Information Technology, \\The University of New South Wales, Canberra, ACT 2610, Australia}	
% 	\IEEEauthorblockA{raselbdeee@gmail.com,}
% }
 
\author{\IEEEauthorblockN{Md Rasel Sarkar\IEEEauthorrefmark{1},
Sreenatha G. Anavatti\IEEEauthorrefmark{1},
Tanmoy Dam\IEEEauthorrefmark{2}, 
Mahardhika Pratama\IEEEauthorrefmark{3}, and
 Berlian Al Kindhi\IEEEauthorrefmark{4}}
 
\IEEEauthorblockA{\IEEEauthorrefmark{1}School of Engineering \& Information Technology, The University of New South Wales, Canberra, ACT 2610, Australia}
\IEEEauthorblockA{\IEEEauthorrefmark{2}Saab-NTU Joint Lab, Nanyang Technological University , 637460, Singapore}
\IEEEauthorblockA{\IEEEauthorrefmark{3}
STEM, University of South Australia, Adelaide 5095, Australia}
\IEEEauthorblockA{\IEEEauthorrefmark{4}Department of Electrical Automation Engineering, Institut Teknologi Sepuluh Nopember, Surabaya, Indonesia}
}

\maketitle

\begin{abstract}
%Time series forecasting uses in everyday life to assist individuals in managing resources and making choices. Therefore, enhanced wind power forecasting (EWPF) is designed for handling power grid operations and boosting power market competition. As a result, reliable large-scale integration of wind power relies in large part on accurate wind power forecasting (WPF). Models including recurrent neural network (RNN), long short-term memory (LSTM), and gated recurrent unit (GRU) have been used for a few years to predict wind power. To efficiently obtain correlations of long-range time-series data sequences under investigation, a more complex and adaptive model is necessary. Transformers have the potential to effectively forecast future values in time-series data, particularly when there are long-term trends or seasonal patterns present. This is due to their ability to surpass the limitations of recurrent neural networks in natural language processing (NLP). This paper proposes an EWPF transformer model for WPF, which overcomes the limitations of RNNs and offers improved time-series forecasting effectiveness. The proposed model is evaluated for single-step and multi-step WPF and compared with GRU and LSTM models on a wind power dataset. The findings of this study have implications for energy producers and researchers in the field of WPF.

%Time series forecasting uses in everyday life to assist individuals in managing resources and making choices. Therefore, 
The main objective of this study is to propose an enhanced wind power forecasting (EWPF)  transformer model for handling power grid operations and boosting power market competition. It helps reliable large-scale integration of wind power relies in large part on accurate wind power forecasting (WPF). 
%Models including recurrent neural network (RNN), long short-term memory (LSTM), and gated recurrent unit (GRU) have been used for a few years to predict wind power. To efficiently obtain correlations of long-range time-series data sequences under investigation, a more complex and adaptive model is necessary. Transformers have the potential to effectively forecast future values in time-series data, particularly when there are long-term trends or seasonal patterns present. This is due to their ability to surpass the limitations of recurrent neural networks in natural language processing (NLP). 
%This paper proposes an EWPF transformer model for WPF, which overcomes the limitations of RNNs and offers improved time-series forecasting effectiveness. 
The proposed model is evaluated for single-step and multi-step WPF, and compared with gated recurrent unit (GRU) and long short-term memory (LSTM) models on a wind power dataset. 
The results of the study indicate that the proposed EWPF transformer model outperforms conventional recurrent neural network (RNN) models in terms of time-series forecasting accuracy. In particular, the results reveal a minimum performance improvement of 5\% and a maximum of 20\% compared to LSTM and GRU. These results indicate that the EWPF transformer model provides a promising alternative for wind power forecasting and has the potential to significantly improve the precision of WPF. The findings of this study have implications for energy producers and researchers in the field of WPF. 
\end{abstract}


\section{Introduction}
Renewable energy has come into prominence as a possible alternative energy source due to the declining supply of fossil fuels and the continued increase in power consumption \cite{scarabaggio2021distributed}. Wind power is an efficient, renewable, and environmentally friendly option for solving the issue of fossil fuel degradation because of its cheap expenses for operation, environmental friendliness, and capacity to produce energy on a commercial scale. Conversely, the reliability of power system operation is reduced by the variability of wind speed, indirectness, and the low density of energy \cite{dong2021wind}\cite{sarkar2019comparative}. In order to ensure the safe and cost-effective functioning of power grids with large-scale wind power, enhancing the accuracy of wind power forecasts has become a critical issue for the existing power system \cite{ma2022hybrid}. There are three ways to predict wind power so far: the physical model, the statistical model, and the machine learning model. Physical modeling approaches attempt to provide a realistic mathematical representation of WPF using data on geography and climate. Realistic real-time prediction tasks may not benefit from this approach as the unstable nature involved could make it impractical \cite{zhao2016improved, daminterval,dam2014ts}. However, statistical techniques use error minimization to determine the ideal correlation between projected wind power and past information \cite{wang2017deep}. Conventional statistical time series forecasting (TSF) methods, namely state space models (SSMs) \cite{durbin2012time}, autoregressive (AR) \cite{meek2002autoregressive},  and moving average (MA) \cite{marshall2017time}, and autoregressive integrated moving average (ARIMA) \cite{brockwell2002introduction} models are fitted to each time series individually \cite{dam2015block, dam2017clustering}. Statistical models are especially well-suited for situations when the series data structure is well understood since they explicitly integrate structural assumptions in the model \cite{daminterval}. The most common challenges of statistical models are;  weaker performance in making long-term predictions, not being suitable for time series that depend on the seasons, and doubts about the reliability \cite{gomes2012wind}. These models make interpretable but require some work to identify the structure. However, deep neural networks (DNNs) \cite{salinas2020deepar, le2019shape} are one example of a kind of machine learning model that  has been offered as an alternate way to address the aforementioned challenges, with RNNs \cite{lai2018modeling, smith2014evolutionary} being used to represent time series in an autoregressive form. When dealing with sequential data, DNNs approaches such as  recurrent models are often regarded as among the most cutting-edge methods currently available. In order to determine the probability of future occurrences, RNN-based models like LSTM \cite{hochreiter1997long} and GRU \cite{cho2014properties} are used. These models allow for the extraction of lengthy dependencies from past sequences. Elsewhere, conventional RNNs are challenging to train because of concerns with vanishing and exploding gradients, which cause the network to be unstable and unable to acquire the knowledge links among input items from enormous amounts of sequential information. The vanishing gradient issue was resolved using techniques like LSTM and GRU despite the advent of recurrent variants. Besides, there are some major challenges that still exist. Since the model is  processing the data sequentially, it is unable to be trained in parallel and unable to capture long-range dependencies. In addition, it shows  poorer performance for non-stationary time series datasets and training needs an input of a fixed size \cite{khandelwal2018sharp} \cite{webster1992tokenization}.

Transformers have obtained superior performance in a variety of natural language processing (NLP), image processing, and other tasks, which has piqued the interest of the community of time series researchers. In recent years, transformer architecture has gained significant recognition in recent years as an advanced method that can surpass conventional RNN-based networks through its ability to acquire global information using an attention mechanism. \cite{lin2020springnet}. Moreover, before using these models, they have to pick a cut-off length. This makes the choice of the cut-off length random as well as subjective. In numerous real-world applications, the essential information in an extended sequence of data is concealed locally and globally. However, unlike time series data, the quadratic model does not exhibit the same growth patterns. Due to the quadratic difficulty of sequence length, RNNs are computationally expensive, unable to capture long-dependence time series data, and unable to process input at once. While transformers have many applications their ability to describe time series is particularly promising because of their capacity to capture long-range relationships and interactions \cite{wen2022transformers}.

%Overall the literature has identified that the recurrence-based model is limited in handling long-term dependencies, parallel processing, and the inclusion of exogenous variables.
Overall the literature has identified that a more complex and adaptive model is required to obtain efficient correlations of long-range time-series data sequences. 
Therefore, it is aimed to evaluate the utility of the transformer model in WPF and to perform detailed comparisons with GRU and LSTM for single-step and multi-step. The proposed transformer model has overcome the recurrent issues of recurrent models. In addition, the model can be captured the dependencies of long-term  wind power for forecasting. Finally, The proposed model has been carried out with LSTM and GRU to verify the improved performance of the proposed method.
%The results indicate that the transformer-based model captures long-range dependencies significantly better than either the LSTM or GRU network-based models. The presented multi-head attention-based transformer model is made up of eight heads and a stack of six identical encoder and decoder layers.

The structure of this paper is as follows: Section 2 presents the problem formulation. The design of the EWPF transformer model is described in detail in Section 3. In addition, the experimental details are shown in Section 4. Section 5 elaborates on the results and discussion. Lastly, Section 5 concludes this paper.

%\section{Attention based Transformer Learning}

\section{problem formulation}

The objective of time-series forecasting is to predict future values based on past observations. Given a time-series X with length T, represented as $X = [x_1, x_2, ..., x_T]$, where $x_t$ is a part of the time series at time instant $t$-th, the problem is to use a model $\omega$ to make step-ahead predictions for $f_{\omega}$, parameterized by $\omega$, that maps the input space to the target space, $f_{\omega}: X \rightarrow Y$. Each input, $x_t$, and target, $y_t$, belongs to $\mathbb{R}^{\mathbb{D}+1}$, where $\mathbb{D}$ is the dimension of feature where the feature is a univariate at the $t$-th time index. The model $f_{\omega}(.)$ is represented by wind power forecasting using a transformer. The target attribute, $y_t$, is set equal to $x_{t+\tau}$ where $\tau$ is the number of time steps. In this paper, both single-step and multi-step time-series forecasting are taken into consideration, with predictions made well ahead of time where $\tau$ is large.

\section{Design of EWPF Transformer}

\begin{figure*}
    \centering
    \includegraphics[scale=0.6]{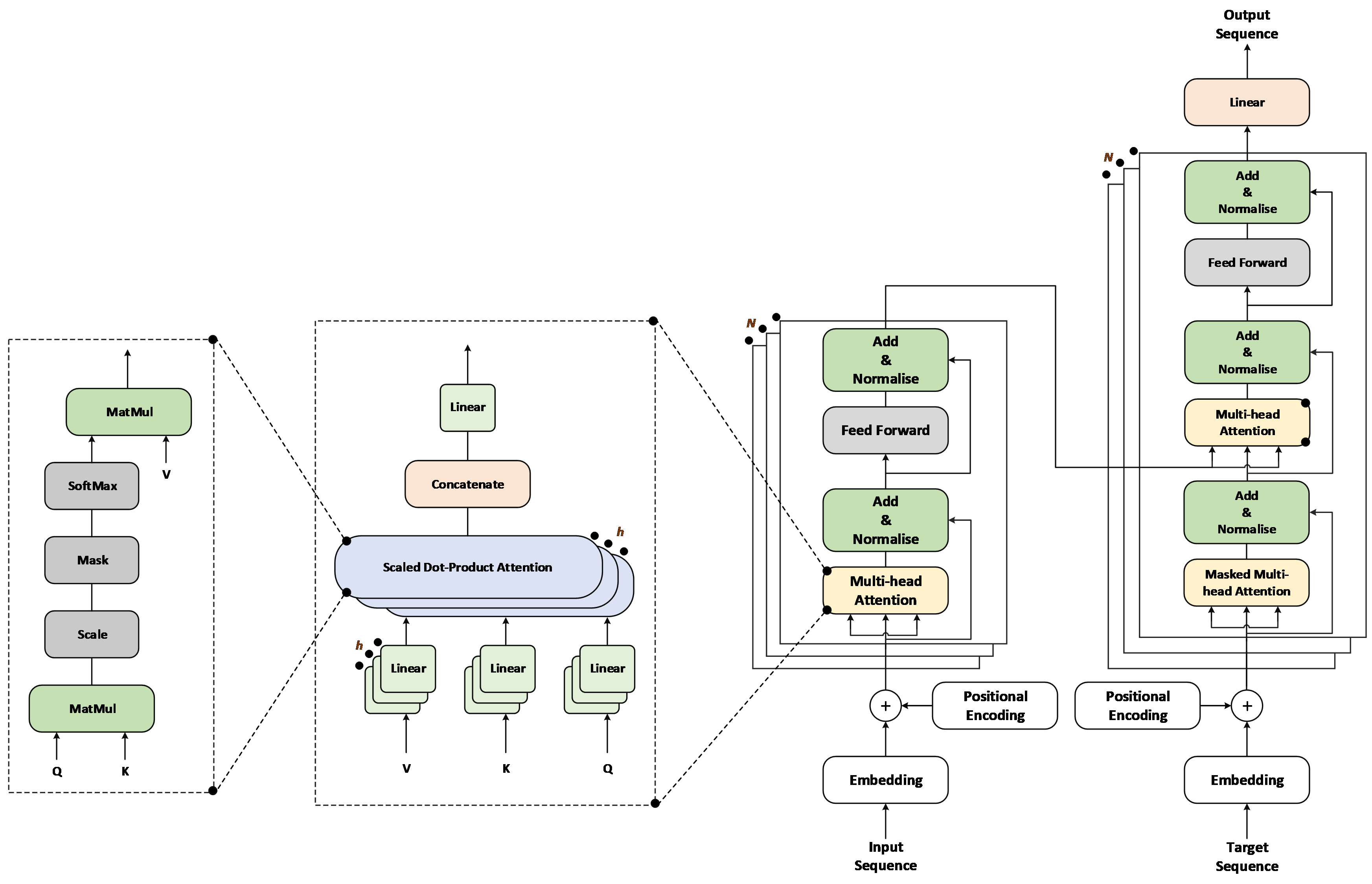}
    \caption{ Architecture of transformer of enhanced wind power forecasting \cite{vaswani2017attention}.}
    \label{fig:f1}
\end{figure*}

The transformer architecture is a novel approach to sequence data processing that was first proposed by Vaswani et al. in 2017 \cite{vaswani2017attention}. Since then, it has been widely used for  NLP tasks \cite{raffel2019exploring, mullick2019d}, and one of its main advantages is its ability to process entire sequences of data \cite{wu2020deep}, \cite{li2019enhancing}. Although the transformer was initially designed for NLP, recent studies have explored its application to time series data. Li et al. \cite{li2019enhancing} compared the transformer with the ARIMA method and RNN-based models, finding that the transformer had superior performance. A multi-horizon univariate time series was forecasted using the transformer architecture by Lim, Arik, et al. \cite{lim2021temporal}. In the task of predicting influenza prevalence, Wu, Green et al. \cite{wu2020deep} employed a transformer architecture and demonstrated its superiority over ARIMA, LSTM, and GRU sequence-to-sequence models with attention. Although the transformer model was modified to suit the time series data, its decoder component was not included. In terms of accuracy, this transformer model achieved a score of 0.738 ($R^2$), while an LSTM model scored 0.4868 ($R^2$), with $R^2$ representing the coefficient of determination. Song, Rajam et al.\cite{song2018attend} utilized the transformer to forecast continuous values for four distinct clinical datasets that consisted of multivariate time series data. Contrariwise, the decoder part of the model was eliminated, and an encoder block was employed to process the multivariate time series data. Subsequently, a regression block was integrated into the model to generate the continuous value predictions. The study found that an LSTM model performed similarly to the transformer model in terms of accuracy for one of the datasets. Nevertheless, considering the transformer architecture's superior performance in handling multivariate time series data over other neural networks, the model utilized in this study was a modified version of the transformer architecture introduced by Zerveas et al. \cite{zerveas2021transformer} and N. Wu et al. \cite{wu2020deep}. To the best of my knowledge, there has been the minimal exploration of EWPF transformers utilizing multi-head attention-based models for single and multi-step forecasting, despite the considerable interest in this area. Fig. \ref{fig:f1} illustrates the details of the model.

\subsection{Input Embedding}
To begin with, the time series data undergoes embedding. Since time is a crucial factor in time series problems, unlike other types of sequential data, the model must take it into account. The transformer model does not have any recurrence or convolution, and as a result, it must rely on the sequence order. Thus, to enable the model to utilize the sequence order, information about the relative or absolute location of the tokens in the sequence needs to be incorporated. This is accomplished by employing positional encoding for the input embedding at the encoder and decoder stack bottoms. The positional encodings and embeddings have the same dimension $d_{model}$, and the formula for the input embedding is as follows \cite{gehring2017convolutional}.

\begin{equation}
   Emd_{\gamma} = PE_{(pos,2t)}+  PE_{(pos,2t+1)}
\end{equation}
where, 
\begin{equation}
   PE_{(pos,2t)} = sin (pos/10000^{2i/d_{model}})
\end{equation}

\begin{equation}
   PE_{(pos,2t+1)} = cos (pos/10000^{2i/d_{model}})
\end{equation}

where $i$ is the individual dimension of the embedding and \textit{pos} denotes the position of the index of the input sequence. Specifically, each positional encoding \textit{sinusoid} functions correspond to even positions, while \textit{cosine} functions correspond to odd positions.

\subsection{Encoder}
The transformer encoder comprises several blocks denoted as ${E_{\theta}^b}$, with $b$ representing the block number. Each block is divided into two sub-layers: the first one is a multi-head self-attention mechanism, while the second is a simple fully-connected position-wise feed-forward neural network (FFNN). To compute the output of each sub-layer, residual connections with layer normalization are utilized \cite{ba2016layer}, \cite{he2016deep}. To be specific, the output is obtained through the expression $LayerNorm (x + Sublayer(x))$, where $Sublayer(x)$ represents the function of the sub-layer. To support these residual connections, all sub-layers and embedding layers in the model have a uniform dimension $d_{model} = 512$, which is also employed in the transformer model discussed in \cite{vaswani2017attention}. The $d_{model}$ corresponds to the number of dimensions of the model used for input to the multi-head attention mechanism. The term "attention function" refers to a mapping between a query and a set of key-value pairs, where both the query and the key-value pairs are vectors. The output is computed as a weighted sum of the values, where the weight assigned to each value is determined by the query's compatibility function with the corresponding key.

\subsubsection{Scaled Dot-Product Attention}
Special attention in the multi-head attention is namely \textit{‘Scaled Dot-Product Attention’} shown in Fig. \ref{fig:f1}. The input consists of dimension $d_k$ queries and keys, as well as dimension $d_v$ values. The attention can be calculated by the query’s dot product with all keys, dividing each by  $\sqrt(d_k)$, then using a softmax function to get the weights on the values. In execution, it computes the attention function on a set of queries at the same time, which are grouped into a matrix Q. The matrices K and V also contain a dense packing of the keys and values. The attention constructs the output matrix as follows:

\begin{equation}
  Attention(Q,K,V)=softmax (\frac{QK^{TM}}{\sqrt{d_k}})V  
\end{equation}

In addition, both additive attention and dot-product (multiplicative) attention \cite{bahdanau2014neural} are widely utilized. However, Dot-product attention with a scaling factor of $1/\sqrt{(d_k)}$ is used in the proposed system. Specifically, a single hidden layer FFNN is used for the determination of the compatibility function with the help of additive attention. Dot-product attention has the same theoretical complexity as the alternative technique, but it can be carried out with optimized matrix multiplication, which makes it more time- and space-efficient. Although both strategies perform fairly well for small amounts of $d_k$, when applied to bigger values of $d_k$, the additive attention strategy outperforms the dot-product attention strategy without any scaling \cite{britz2017massive}.

\subsubsection{Multi-head Attention}
Instead of using a single attention function with keys, queries, and values of dimension $d_{model}$, it was discovered that projecting the queries, keys, and values times with learned linear projections to dimensions $d_q$, $d_k$, and $d_v$, respectively, was more beneficial. This resulted in head ($h$) parallel versions of queries, keys, and values that could be processed by the attention mechanism. The output values had a dimension of $d_v$ and were combined and projected again to obtain the final values. Multi-head attention allows the model to attend to input from different representation subspaces at various stages and averaging prevents over-attentiveness compared to a single-attention head.

\begin{equation}
  MultiHead(Q,K,V)=Concat(head_1.......,head_h)
\end{equation}

where $head_i = Attention (QW^{Q}_{i},KW^{K}_{i}, VW^{V}_{i})$

the projections are parameter matrices and symbolized as $W^{Q}_{i} \in  {\rm I\!R}^{d_{model}\times d_k}$\, $W^{k}_{t} \in  {\rm I\!R}^{d_{model}\times d_k}$\, $W^{V}_{t} \in  {\rm I\!R}^{d_{model}\times d_v}$\,$W^{0} \in  {\rm I\!R}^{h d_{model}\times d_k}$\ where, $W^Q$,$W^K$ and $W^V$ are quires, keys and values weights respectively. There are $h_{head}$ = 8 parallel attention has been used in the multi-head attention layer. In addition, each head has $d_k$ = $d_v$ = $d_{model}/h$ = 64 dimension. The overall computing cost is comparable to that of single-head attention with full dimensions because of the lower dimension of each head.

\subsection{Feed-Forward Neural Network (FFNN)}

Besides the attention sub-layers, the encoder and decoder layers also contain a FFNN that is connected fully and operates on each position in the sequence uniformly and independently. This network comprises two linear transformations separated by a \textit{ReLU} activation function. 
\begin{equation}\label{eq1}
 FFN(x)  = max(0,xW_1+ b_1 )W_2+ b_2
\end{equation}

Although the linear transformations are identical across all positions, the parameters differ from layer to layer. This is analogous to performing two convolutions with kernel size 1. The dimensions of the input and output are both $d_{model}$ = 512, while the inner-layer of the FFNN  has a dimensionality of $d_{ff}$ = 2048. In the transformer model, the input and output embeddings are converted to the input and output vector dimensions of $d_{model}$ = 512. The decoder output is then passed through the next probability model. The transformer model uses the same weight matrix for the two embedding layers and the pre-softmax linear transformation, as proposed by Press and Wolf \cite{press2016using}. The weights in the embedding layers are multiplied by $\sqrt{(d_{model})}$.

\subsection{Decoder}

The transformer model is composed of multiple decoder blocks ${D_{\beta}^b}$, each indexed by $b$. In addition to the two sub-layers present in each encoder layer, a third sub-layer is added in each decoder block, which performs multi-head attention over the encoder stack's output. Like the encoder, each sub-layer in the decoder is surrounded by residual connections and followed by layer normalization. Furthermore, the self-attention sub-layer in the decoder stack is modified to prevent positions from attending to subsequent positions. Along with the output embedding being moved by one position, makes sure that forecasts for position $i$ are only based on known outputs at places that come before $i$. \cite{vaswani2017attention}.  

The final loss function is defined as follows,

\begin{equation}\label{MSEloss}
\mathcal{L}_{MSE}=\sum_{t=1}^{T}\frac{1}{2}(y_t - D_{\beta}((E_{\theta}(x_t)))^2 
\end{equation}

Algorithm \ref{algo_:EWPFT} contains a detailed and thorough description of the EWPF transformer, providing all the necessary information and steps required to understand and implement the algorithm effectively.

\begin{algorithm}[h]
\begin{algorithmic}[1]
\STATE \textbf{Input:} Time series data $X^{train} = \{(x_1,y_1), (x_2,y_2), ..., (x_T, y_{T+\tau})\}$, Similarly, 
$X^{test} = \{(x^t_1,y^t_1), (x^t_2,y^t_2), ..., (x^t_T, y^t_{T+\tau})\}$, number of steps ahead $m$, model parameters, $\{f(.)= E_{\theta}, D_{\beta}, Emd_{\gamma}\}$, batch=$B$, epoch =$n_{epoch}$, Adam hyper$-$
parameters ($l_r , \pi_{1}, \pi_{2}$), time-step=$\tau$
\STATE /* Initialisation of three model component parameters  */
\STATE $Emd_{\gamma}$, $E_{\theta}$ and $D_{\beta}$ $ \sim \mathcal{N}(0, 0.02^2)$ 
\FOR{$p$ in  $\{ 1, 2, ..., n_{epoch}\}$}
\STATE /* Embedding the input with batch, $B \in X^{train}$   */
\STATE  $\{x_i\} _{i=1}^{B} \leftarrow$ $Emd_{\gamma}(\{x_i\} _{i=1}^{B})$
\STATE /* Encoding the embedding output */
\STATE $\{z_i\} _{i=1}^{B} \leftarrow$ $E_{\theta}(\{x_i\} _{i=1}^{B})$ 
\STATE /* Decoding the encode output */
\STATE  $\hat{y}_{i} \leftarrow$ $D_{\beta}\{z_i\} _{i=1}^{B}$
\STATE /* Calculate the MSE loss in (\ref{MSEloss}) */
\STATE /* all the parameters are updated simultaneously */

\STATE $Emd_{\gamma}$ update through MSE loss \\
$\gamma\rightarrow (\nabla_{\gamma}\frac{1}{B}\mathop{\mathcal{L}_{MSE}}),\gamma,  l_{r},\pi_{1},\pi_{2})$
\STATE $E_{\theta}$ update through MSE loss \\
$\theta\rightarrow (\nabla_{\theta}\frac{1}{B}\mathop{\mathcal{L}_{MSE}}),\theta,  l_{r},\pi_{1},\pi_{2})$
\STATE $D_{\beta}$ update through MSE loss \\
$\beta\rightarrow (\nabla_{\beta}\frac{1}{B}\mathop{\mathcal{L}_{MSE}}),\beta,  l_{r},\pi_{1},\pi_{2})$

\IF{$n_{epoch} \mathbin{\%} 1== 0$}
    \STATE /* Estimate of MSE error values */
    \STATE  $f(.)$ for $X^{test}$ datasets.
\ENDIF
\ENDFOR

\end{algorithmic}
\caption{EWPFT}
\label{algo_:EWPFT}
\end{algorithm}

\newpage
\section{ Experimental Details}

In this paper, there are three deep learning models namely, multi-head attention-based transformer, LSTM, and GRU have been investigated. Our model is constructed based on PyTorch with Python 3.7. The hardware specification of the computer is a single NVIDIA Corporation / Mesa Intel® UHD Graphics (TGL GT1) for training. Prior to being converted to sequences, the raw input data is first normalized using a $Min-Max$ $Normalization$ $(-1.0, +1.0)$ technique. Normalisation is a crucial data preprocessing phase in deep learning, as it can enhance the model's efficiency and convergence during training. Normalization entails adjusting the input data so that its characteristics have comparable ranges and distributions, which may contribute to the model's learning. As a result of these findings, we have explored how DNN models do when confronted with varying sequence lengths. There are three distinct representations of the input data sequence, such as 10, 20, and 50. The historical data was associated with the model's input sequences. 

Since there is no recurrent  network layer built into the  transformer encoder, the transformer injects positional information into the embedding using positional encoding. In the position encoding, for each odd and even time step, the sine and cosine functions produced two different vectors. The positional information would be injected by including each of them in the appropriate embedding vector. The self-attention mechanism in the transformer model allows the model to learn relationships between different time steps in the time series. Self-attention computes a weighted sum of all the time steps in the time series, where the weights are learned during training. In addition, the model can assign more weight to relevant time steps and less weight to irrelevant time steps, allowing it to attend to the most important parts of the time series. Moreover, multi-head attention computes multiple attention mechanisms in parallel. This allows the model to capture different types of dependencies and relationships within the time series. Since the transformer model does not use recurrent connections, it needs a way to encode the position of each time step in the time series. This is achieved using positional encoding, which adds a fixed-length vector to the input of each self-attention layer to represent the position of each time step. After the self-attention layers, the output of the transformer model is passed through a series of fully connected layers (FFNN). These layers allow the model to learn complex non-linear relationships between the input time series and the output predictions. The output layer of the transformer model for time series forecasting uses a mask to prevent the model from predicting future time steps that have not occurred yet. This is important because, in time series forecasting, the model should only have access to information that is available at the time of prediction. The training is carried out using the rolling forecasting option (stride size=1).  The encoder and decoder input lengths are 10, 20, and 50, respectively, which correspond to the prediction lengths of the $1^{st}$ and  $5^{th}$  steps. We update the dropout rate for the progressive training scheme every 5 iterations, with a dropout limit of 0.1. Moreover, during training, the model is optimized to minimize the difference between the predicted values and the true values of the target time series. This is done using a loss function such as mean squared error (MSE) or mean absolute error (MAE). The transformer for time series forecasting is a robust model that can capture long-term dependencies and relationships within a time series, making it well-suited for accurate and efficient forecasting. Table 1 contains more information about the model parameters. To evaluate the performance of the models, two metrics are used namely MSE and MAE, and $R^2$.

 A transformer is built with a stack  encoder and decoder with self-attention layers that were connected point-by-point. It mapped an input sequence. The experimental settings for the  transformer's  encoder and decoder block are as follows: $d_{model}$= $Number_{hidden}$ = 512, $Number_{layer}$ = 8, and $dropout$ = 0.1, $Input_{feature}$ 
 and $Output_{feature}$ = 1. In addition, loss and optimization were calculated using the MSE loss function and adaptive moment estimation (Adam) \cite{boonlia2022improving} with a $l_{r}$ of 0.0001 \cite{dam2020mixture, dam2022latent} for transformer, LSTM, and GRU.  Hence, the aim of the paper is to identify the best model from the transformer, LSTM, and GRU for time series forecasting. Therefore, the same parameters have been set for model evaluation. The LSTM and GRU model parameters are $Number_{hidden}$=512, $l_{r}$ = 0.0001, $\pi_{1}=0$, $\pi_{2}=0.5$ \cite{dam2021does}, $Number_{layer}$ = 8.

\begin{table}
	\centering
	\caption{crucial variables of the Transformer}
	\label{tab:T}
	\begin{tabular}{ccc}
		\hline \noalign{\smallskip}
		\textbf{Category} & \textbf{Hyper-parameter} & \textbf{Parameter Value} \\ \hline
		\multirow{6}{*}{Structure of Model}   & $d_{model}$ &  512  \\
		& $Number_{hidden}$ & 512  \\
		& $Number_{head}$ & 8  \\
		& $Number_{layer}$ & 8  \\
		& $Kernel_{size}$   & 1  \\
		   \hline \noalign{\smallskip}
		\multirow{6}{*}{Hyper-parameters}   & dropout &  0.1  \\
		& Optimizer & Adam  \\
		& Learning rate $(l_r)$ & 0.0001  \\
		& Batch Size & 64  \\
		\hline  		
	\end{tabular}
\end{table}

\begin{figure*}[]
\begin{minipage}[t]{0.5\linewidth}
    \centering

    \includegraphics[width=1\textwidth]{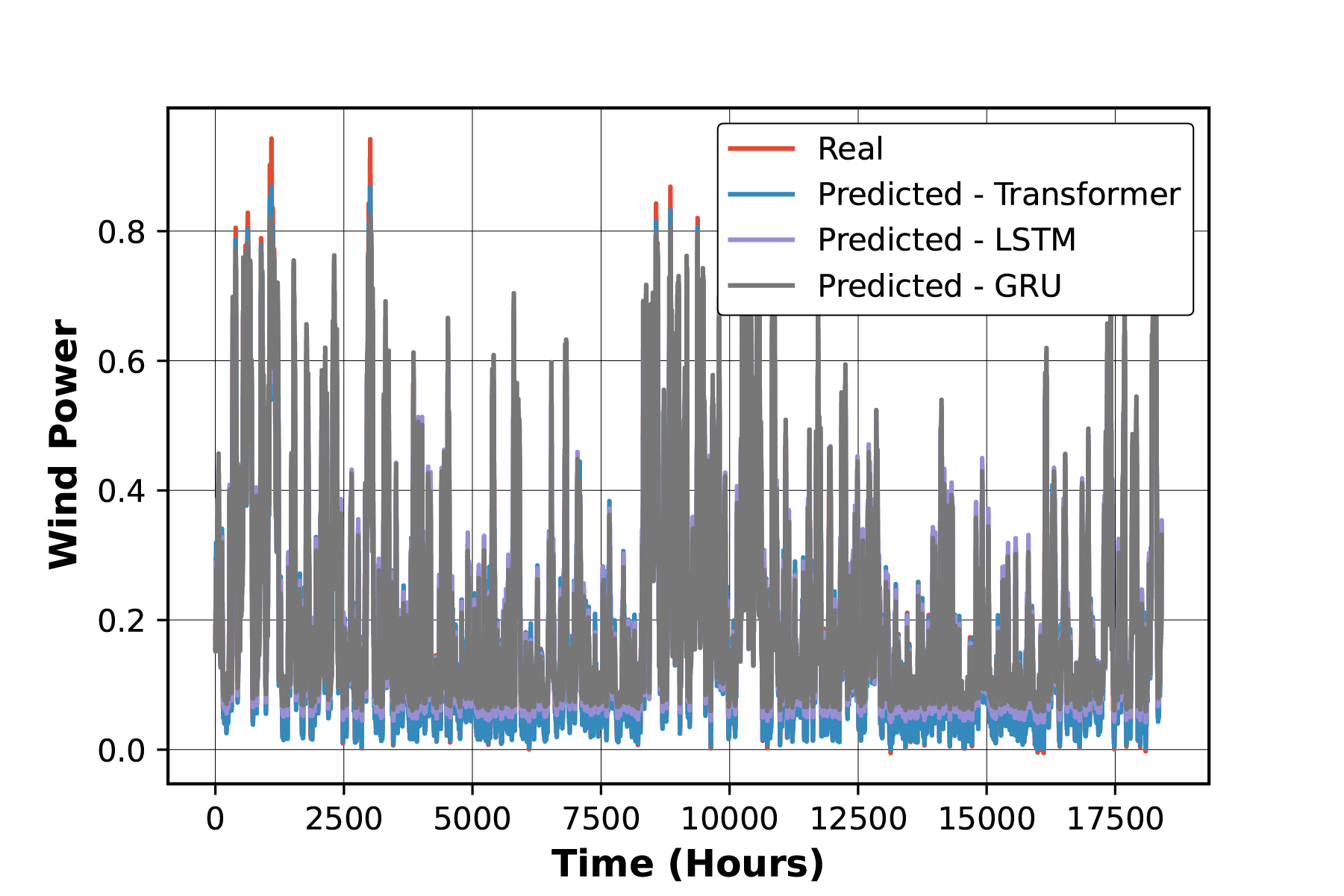}
    \caption{Single-step wind power forecasting for 20 sequences.}
    \label{fig:f2}
\end{minipage}
\hspace{0.1cm}
\begin{minipage}[t]{0.5\linewidth} 
    \centering
    
    \includegraphics[width=1\textwidth]{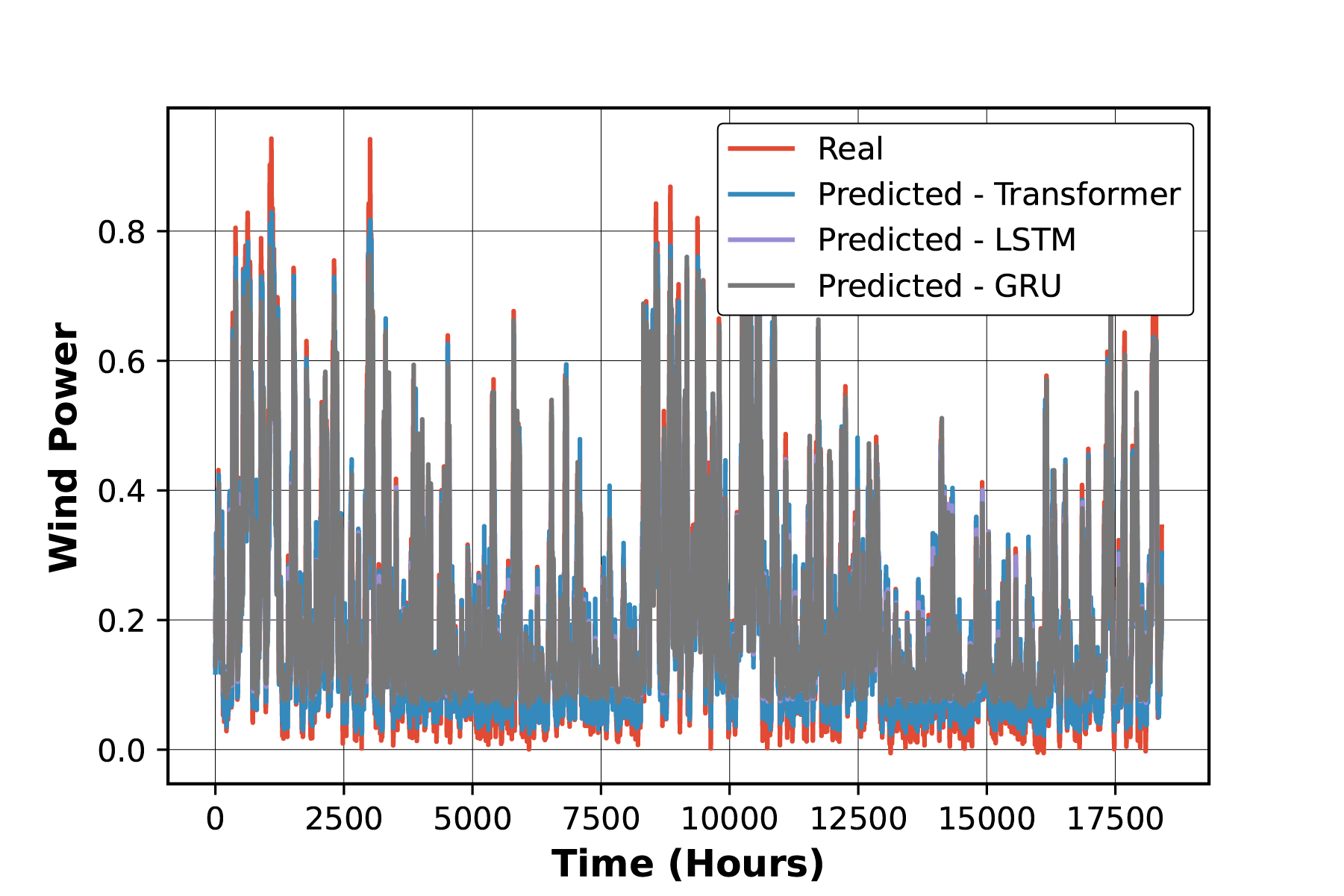}
    \caption{Multi-step wind power forecasting for 20 sequences.}
    \label{fig:f3}
\end{minipage}        
\end{figure*} 

\section{Results and Discussion}
\begin{figure*}
\begin{minipage}[t]{0.5\linewidth}
    \centering
    \includegraphics[width=1\textwidth]{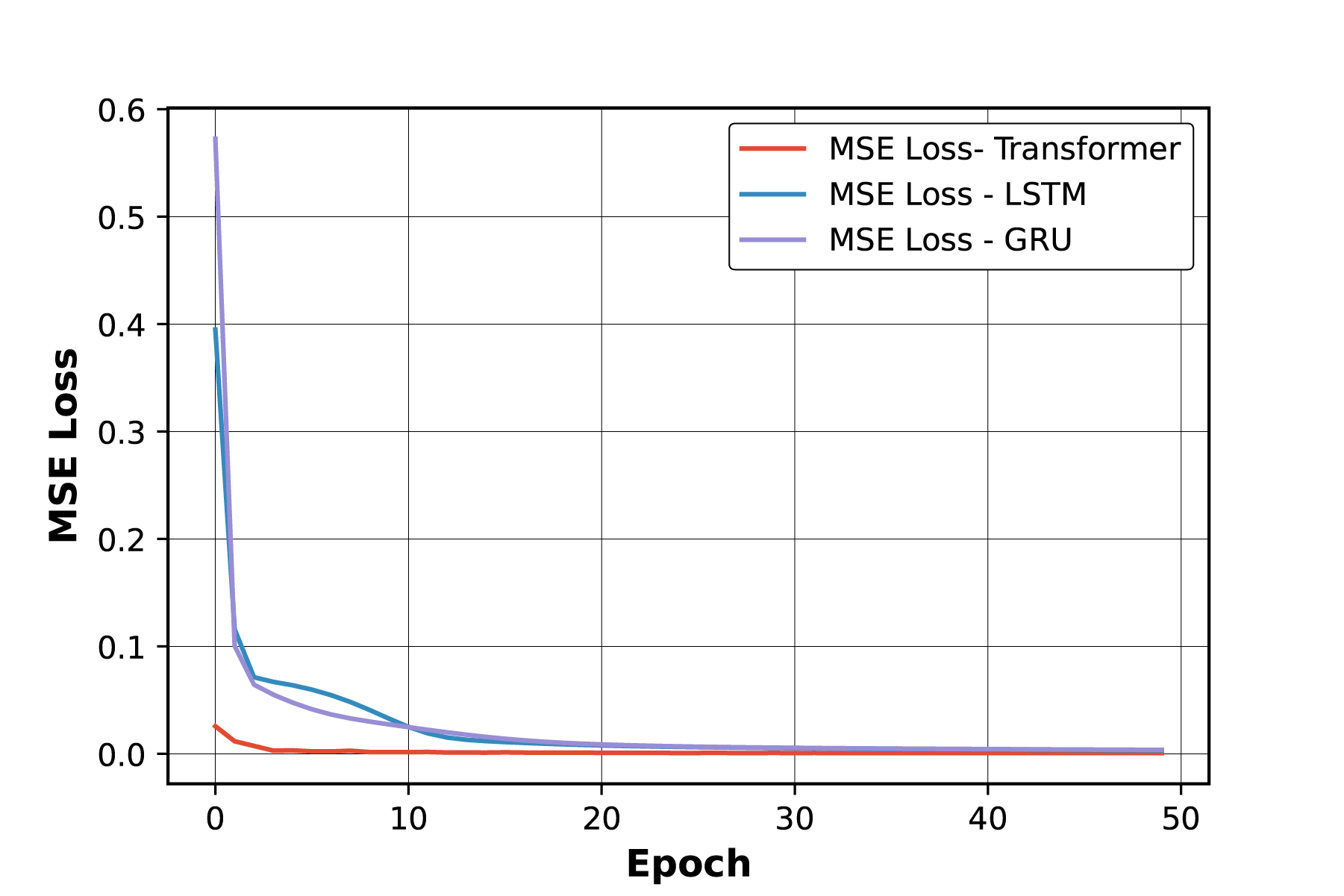}
    \caption{MSE loss of single-step WPF for 20 sequences.}
    \label{fig:f4}
\end{minipage}
\hspace{0.1cm}
\begin{minipage}[t]{0.5\linewidth} 
    \centering
    
    \includegraphics[width=1\textwidth]{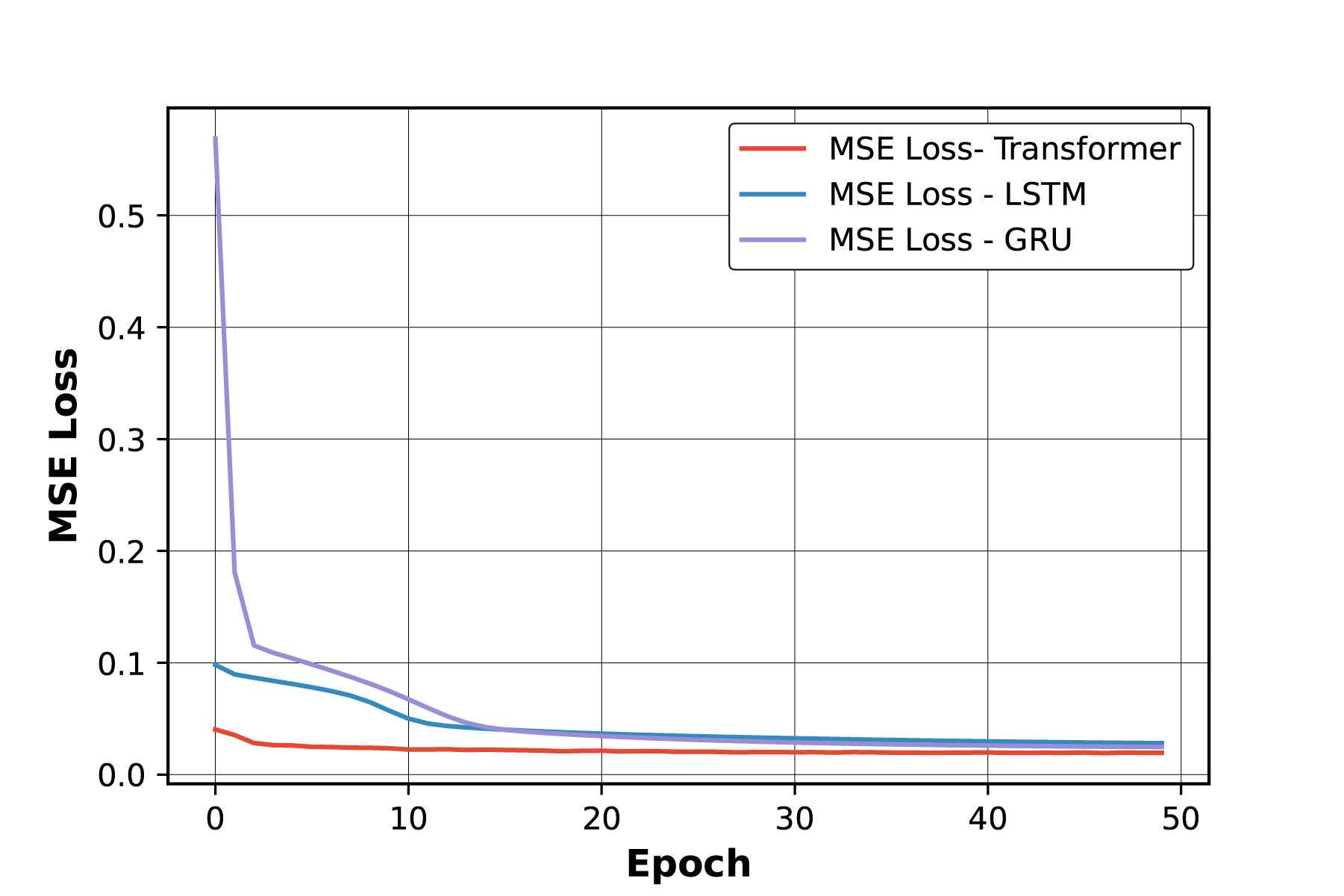}
    \caption{MSE loss of multi-step WPF for 20 sequences.}
    \label{fig:f5}
\end{minipage}        
\end{figure*}

Wind farms depend on wind power predictions. Accurate forecasting is needed to keep the electrical grid stable as it is interrupted and varied. Grid operators need wind power forecasts to balance energy supply and demand in real time. Moreover, single-step and multi-step forecasts both look into the future and estimate values, but they differ in key ways. The single-step method trains a model to make predictions just one step into the future. Whereas, there are no long-term trends or seasonal patterns, and the values of the time series in the future rely simply on their values in the past, this method may be helpful. In contrast, the model is educated to anticipate a certain number of steps rather than a single eventual state. Meanwhile, there are long-term trends or seasonal patterns in the data, and the future values of the time series rely on both the past values and the future values, multi-step prediction is helpful. However, two comparative experiments were carried out in this paper to validate the robustness of the proposed model for single and multi-step WPF. Three models—transformer, LSTM, and GRU are used to process a series of data in order to undertake single-step (1-hour) and multi-step (5-hour) wind power forecasts in order to evaluate model effectiveness. In the experiment, the proposed model utilized an open historical dataset from Germany, which included actual measurements of wind power (Watt) generation. Each record in the dataset included one parameter sampled every hour, for a total of 61,500 samples. In addition, it is divided into training $(70\%)$ and testing $(30\%)$. The data has been divided into 10, 20, and 50 input sequences. These sequences are applied in the LSTM, GRU, and transformer models for both single-step and multi-step forecasting. In a single step, the next step of the input sequence is forecasted and in a multi-step, the $5^{th}$ step of the input sequence is forecasted.  After several trial errors using different epochs, the epochs with 50 setups were used for the training of the model. After 50 epochs, there was no longer any variation in the MSE, thus that number was used. In the result section, only 20 input sequence results are shown for single and multi-step. Fig. \ref{fig:f2} shows the single-step time series forecasting for 20 sequences of input data. Fig. \ref {fig:f3} shows the multi-step time series forecasting for 20 sequences of input data. In both cases, experimental results show that the transformer  model provides better performance when compared to the most popular deep learning models, LSTM and GRU, respectively. From Fig. \ref{fig:f4} and Fig. \ref{fig:f5}, it can be said that the transformer  model shows a lower error when comparing the LSTM and GRU. This will also be advantageous when forecasting multivariate time series. In addition, in our investigation, three performance indicators are used.

\begin{table}
	\centering
	\caption{Performance indicators of Transformer, LSTM, and GRU.}
	\label{tab:T}
	\begin{tabular}{cccccc}
		\hline \noalign{\smallskip}
		\textbf{Step} & \textbf{Sequence} & \textbf{Model } & \textbf{MSE}& \textbf{MAE} & \textbf{$R^{2}$} \\ \hline
	\multirow{9}{*}{Single-step}   & \multirow{3}{*}{10} & LSTM & 0.0029 & 0.396 & 0.9729  \\
		   && GRU & 0.0038 & 0.552 & 0.9723  \\
		  && \textbf{Transformer} & 0.0010 & 0.095 & 0.9923  \\ \noalign{\smallskip}
		   & \multirow{3}{*}{20} & LSTM & 0.0054 & 0.442 & 0.9672  \\
		  && GRU & 0.0038 & 0.610 & 0.9695  \\
		  && \textbf{Transformer} & 0.0008 & 0.0790 & 0.9940  \\ \noalign{\smallskip}
		   & \multirow{3}{*}{50} & LSTM & 0.0023 & 0.3932 & 0.9701  \\
		  && GRU & 0.0040 & 0.443 & 0.9706  \\
		   && \textbf{Transformer} & 0.0007 & 0.0820 & 0.9958  \\ \hline \noalign{\smallskip}
		   
      \multirow{9}{*}{Multi-step}   & \multirow{3}{*}{10} & LSTM & 0.0345 & 0.856 & 0.7454  \\
		   && GRU & 0.0598 & 1.340 & 0.5622  \\
		   && \textbf{Transformer} & 0.0212 & 0.480 & 0.8331  \\ \noalign{\smallskip}
		   & \multirow{3}{*}{20} & LSTM & 0.345 & 0.962 & 0.7496  \\
		   && GRU & 0.2876 & 0.911 & 0.7812  \\
		   && \textbf{Transformer} & 0.1901 & 0.456 & 0.8471  \\ \noalign{\smallskip}
		   & \multirow{3}{*}{50} & LSTM & 0.294 & 0.845 & 0.7833  \\
		   && GRU & 0.0263 & 0.876 & 0.8067  \\
		   && \textbf{Transformer} & 0.0103 & 0.650 & 0.8431  \\ \hline 
		
	\end{tabular}
\end{table}

Table 2 presents the performance indicators of three different neural network models: LSTM, GRU, and Transformer, for time series prediction tasks. The models were evaluated using three different sequence lengths: 10, 20, and 50, for both single-step and multi-step prediction tasks. The evaluation is divided into two types of prediction tasks: single-step and multi-step prediction. In single-step prediction, the task is to predict the next value in the time series, given the current value. In multi-step prediction, the task is to predict a sequence of future values, given the current value. For each model, sequence length, and prediction task, Table 2 reports three performance indicators: MSE, MAE, and $R^2$.
The MSE measures the average squared difference between the predicted and actual values, and it gives an indication of the accuracy of the model. The MAE measures the average absolute difference between the predicted and actual values, and it indicates how far off the model's predictions are from the actual values. The $R^2$ measures how well the model fits the data, with higher values indicating a better fit.
Looking at the results, the transformer model consistently outperformed the LSTM and GRU models across all sequence lengths and prediction tasks. Specifically, for single-step prediction tasks, the transformer model achieved the lowest MSE and MAE values and the highest $R^2$ coefficients. For the sequence length of 10, the transformer model has an MSE of 0.001, an MAE of 0.095, and an $R^2$ coefficient of 0.992, which are better than the corresponding values for the LSTM and GRU models.
For multi-step prediction tasks, the performance difference between the models is relatively smaller, but the transformer still achieves the lowest MSE and highest $R^2$ coefficient across all sequence lengths. For instance, for the sequence length of 20, the transformer model has an MSE of 0.1901 and an $R^2$ coefficient of 0.8471, which were better than the corresponding values for the LSTM and GRU models. the results suggest that the transformer model is a promising option for time series prediction tasks, as it defeated the LSTM and GRU models in terms of accuracy and fit, for single-step and multi-step prediction tasks.

\section{Conclusion}
Forecasting time series is a difficult endeavor because time series contain both linear and non-linear trends. In this paper, the transformer model has been used to make predictions about time series. The transformer has been shown to be able to capture long-term dependencies in wind power data, but LSTM and GRU may struggle.  The attention-based transformer is used to make wind power forecasting faster and more accurate. To determine the impact of the long sequence input data, the proposed model applies several input sequences, such as 10, 20, and 50 processes, to a multi-head attention-based transformer, LSTM, and GRU. As measurement indices, the MSE, MAE, and $R^2$ were employed. The results showed that the proposed multi-head attention-based transformer model performed more effectively than LSTM and GRU by making the MSE, MAE, and $R^2$ values better for input sequences of 10, 20, and 50 for both single-step and multi-step forecasting. Future efforts will focus on the creation of more efficient and precise hybrid transformer prediction methods for multivariate wind power forecasting. Furthermore, the hybrid transformer will be applied to the irregular time series forecasting model to handle the model's irregularity.

\clearpage
\bibliographystyle{IEEEtran}
\bibliography{bibliography.bib}

\end{document}